\documentclass[12pt]{article}
\input{amssym.def}
\input{amssym}
\usepackage[dvips]{color}
\usepackage{epsfig}
\usepackage{hyperref}
\normalbaselines
\makeatletter
\@addtoreset{equation}{section}
\makeatother

\def\be{\begin{equation}}
\def\ee{\end{equation}}
\def\ba{\begin{eqnarray}}
\def\ea{\end{eqnarray}}

\newcommand{\rf}[1]{(\ref{#1})}

\def\ket#1{|#1\rangle}

\def\pmax{p_{\scriptsize{\mbox{max}}}}
\def\taumax{\tau_{\scriptsize{\mbox{max}}}(L,t)}

\begin{document}

\title
{From conformal invariance to quasistationary states}

\author{Francisco C. Alcaraz$^1$   and 
Vladimir Rittenberg$^{2}$
\\[5mm] {\small\it
$^1$Instituto de F\'{\i}sica de S\~{a}o Carlos, Universidade de S\~{a}o Paulo, Caixa Postal 369, }\\
{\small\it 13560-590, S\~{a}o Carlos, SP, Brazil}\\
{\small\it$^{2}$Physikalisches Institut, Universit\"at Bonn,
  Nussallee 12, 53115 Bonn, Germany}}
\date{\today}
\maketitle
\footnotetext[1]{\tt alcaraz@if.sc.usp.br}
\footnotetext[2]{\tt vladimir@th.physik.uni-bonn.de}

\begin{abstract}
 In a conformal invariant one-dimensional stochastic model, a certain
non-local perturbation takes the system to a new massless phase of a
special kind. The ground-state of the system is an adsorptive state. Part of
the finite-size scaling spectrum of the evolution Hamiltonian stays unchanged
but some levels go exponentially to zero for large lattice sizes becoming 
degenerate with the ground-state. As a consequence one observes the 
appearance of quasistationary states which have a relaxation time which grows
exponentially with the size of the system. Several initial conditions have 
singled out a quasistationary state which has in the finite-size scaling 
limit the same properties as  the stationary state of the conformal 
invariant model.
\end{abstract}

\section{ Introduction} \label{sect1}

We have recently \cite{AP} presented the peak adjusted raise and peel model 
(PARPM). This is a one-parameter (denoted by $p$) extension of the well 
studied, raise and peel (RPM) one-dimensional growth model \cite{GNP,AR3}. The latter 
model is recovered if one takes $p = 1$. The PARPM is
defined in the configuration space of Dyck (RSOS) paths on a lattice with
$L + 1$ sites ($L$ even). The RSOS configurations can be seen as an interface separating a
fluid of tiles covering a substrate and a rarefied gas of tiles hitting the 
interface. If $h(i)$ ($i = 0,1,\ldots,L$) is the height at the site $i$ 
of an RSOS path, for the substrate one has $h(2k) = 0, h(2k+1) = 1$. The interface is 
composed of clusters which touch each other. Depending on the position of the
hit, the tile can be locally adsorbed (increasing the size of a cluster or
fusing two clusters) or can trigger a nonlocal desorption, peeling part of a
layer of tiles from the surface of a cluster. A tile hits a peak 
(local maximum) with a $p$ dependent probability,  and is reflected. The other 
sites are hit with equal probabilities. The effective rates for adsorption 
and desorption become dependent on the total number of peaks of the 
configuration, on the size $L$ of the system, and on $p$.    

 If the parameter $p = 1$, the rates are all equal to 1 and  independent on
the number of peaks and the size $L$ of the system. The situation is very different if $p \neq 1$. 
The dependence of the rates on the global properties of the 
 configuration can be seen as a process with long-range interactions. The
larger the value of the parameter $p$ ($p>1$), the stronger the "long-range" effect is.  Configurations with many peaks become more stable. 
The slowing down of 
 configurations with many peaks will lead us to new physics. 

 It was shown in \cite{AP} that for $0\leq p \leq \pmax $, where 
$\pmax =2(L-1)/L$, in 
the finite-size 
scaling limit, the properties of the system are $p$ independent. The system 
is conformal invariant (this is the merit of the model) and the 
stationary states have many well understood properties. The $p$ 
dependence of the model appears only in the non-universal sound 
velocity $v_s(p)$ which fixes the time scale. The spectra of the Hamiltonians describing the 
time evolution of the system is given by a known representation of the 
Virasoro algebra \cite{SSS}. Moreover it was observed that if 
$p = \pmax$  the
stationary state become an absorbing state, i. e., with 
probability 1 one finds only one configuration. This is a new phase. The 
absorbing state corresponds to the configuration given by the substrate 
(maximum number of peaks). Conformal invariance should be lost. 
 It turns out that the picture is much more complex, the PARPM at
$p = \pmax$ has fascinating properties. One observes quasistationary
states, and conformal invariance is broken in an uncanny way. One should keep
in mind that one deals with a nonlocal model.
For
$p >\pmax $ some rates become negative and the stochastic process is ill 
defined. 
 
 The present paper deals only with the new phase of the PARPM and is a natural continuation
of our previous work \cite{AP}. The presentation of the model in Section 2 in this
 paper is a mere repetition of Section 2 of \cite{AP}.  

 Since except for $p = 1$ the PARPM is not integrable, we have studied its 
properties using Monte Carlo simulations on large systems (up to $L = 70000$). For
the study of the spectrum of the evolution Hamiltonian we have done numerical 
diagonalizations of lattices up to $L = 18$ sites and up to $L = 30$ for one 
special case.

 In Section 3, using Monte Carlo simulations we study the time evolution of 
the system. We show that, surprisingly, 
for moderate system sizes and various initial conditions, after a short 
transient time the system stays practically unchanged 
for a long relaxation time in a quasistationary state (QSS). 
 
 Quasistationary states are observed in systems with long-range or 
mean-field interactions in statistical mechanics and Hamiltonian dynamics. 
There is a long list of papers on this subject and we refer the reader to 
some reviews \cite{CDR,DMU}. Typically the time the system spends in a QQS
increases with the length of the system following a power law (this is not
going to be seen in our case). We are aware of only one other stochastic 
model defined on a lattice (the ABC model \cite{EKKM}) in which QSS's are seen. 
In this model in the stationary state translational invariance is broken. The
sites are occupied in alphabetical order by three blocks of A, B or C particles.  In the quasistationary 
states there are more blocks. As we are going to see our model is very 
different.

 In order to understand the origin of the QSS, in Section 4 
we do a finite-size scaling study of the spectra of the Hamiltonian which 
gives the time evolution of the system. In the unperturbed ($p < \pmax$) 
case, the finite-size scaling spectrum is given by the conformal dimensions
$\Delta$. They are equal to all non-negative integers numbers except 1 (there
is no current). The degeneracies are also known \cite{SSS}. 

 It turns out that in the new phase, all the properly scaled energy levels 
stay unchanged, 
except for one level for each even conformal dimension (this includes the 
energy-momentum tensor which plays a crucial role in conformal 
invariance and has $\Delta = 2$ !). We are thus left with

\be \label{e1.1}
\Delta = 0, 3, 4,...,
\ee
with a degeneracy smaller by one unit for all even values of $\Delta$ 
starting with $\Delta = 4$. This is not a rigorously proven result but a 
conjecture based on numerics up to $\Delta = 7$. 

 If we increase the lattice size of the system, all the levels which are not anymore
in the conformal invariant towers  go exponentially to
zero. This makes the value $\Delta = 0$ to be infinitely degenerate in the
infinite volume limit. 
 One should keep in mind that the configuration corresponding to the
substrate, which is an absorbing state, has also $\Delta = 0$.
For finite volumes, the missing levels are the  origin of the QSS.

 In Section 5, in order to make the connection between eigenfunctions and 
the probability distribution functions (PDF) seen in QSS, we mention some 
properties of intensity matrices (the Hamiltonian is one of them) when one 
of the states is an absorbing state.  
 
 We next derive some properties of the QSS related  to the eigenfunction 
of the  
energy level originally at  $\Delta = 2$ for $p<\pmax$, and which decreases exponentially to
the value $\Delta = 0$ at $p=\pmax$. This correspondence is possible due to a unique 
property of the eigenfunction of the first exited level of an intensity 
matrix in the presence of an absorbing state. 

 Using Monte Carlo simulations and initial conditions in which the 
probability to have the substrate is taken zero, in Section 6, we have 
studied the 
density of contact points and the average height in the finite-size scaling
limit. If the system is conformal invariant, both these quantities are given
by precise analytic expressions. It turns out that in the QSS state,  
 with very high accuracy, the same expressions describe the data. This result is more than surprising.
 As 
 we have discussed, in the new phase the finite size scaling 
spectrum of the Hamiltonian (which gives the time-like correlation 
functions) is not the same as in the conformal invariant region and one 
would expect that the space-like correlation functions to change too.
  
 The open questions and our conclusions are presented in Section 7.

\section{ 
  Description of the peak adjusted raise and peel model}

\begin{figure}
\centering
\includegraphics[angle=0,width=0.5\textwidth] {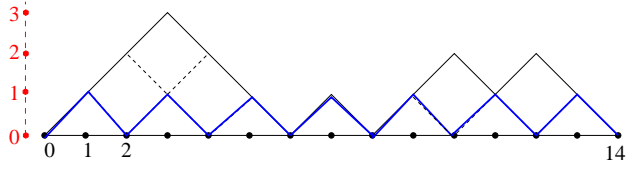}
\caption{
 An example of a Dyck path for L = 14.  There are four contact
points and three clusters. The substrate profile is shown in blue.}
\label{fig1}
\end{figure}

 We consider an open one-dimensional system with $L + 1$
sites ($L$ even). A
Dyck path is a special restricted solid-on-solid (RSOS) configuration  defined as
follows. We attach to each site $i$ non-negative integer heights $h_i$
 which
 obey RSOS rules:
\be \label{e4}
 h_{i+1} - h_i =\pm1, \quad h_0 = h_L = 0 \quad  (i = 0,1,\ldots,L-1).
\ee
There are
\be \label{e5}
Z (L) = L!/(L/2)!(L/2 + 1)!
\ee
configurations of this kind. If $h_j = 0$ at the site $j$ one has a {\it
contact point}.
 Between two consecutive contact points one has a {\it cluster}. There are
four contact points and three clusters in Fig.~\ref{fig1}.

\begin{figure}
\centering
\includegraphics[angle=0,width=0.5\textwidth] {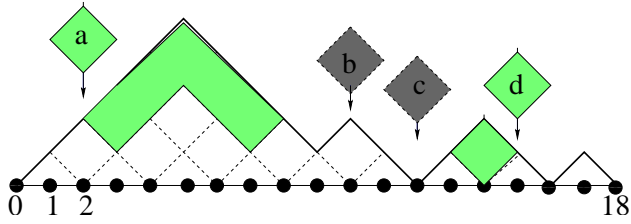}
\caption{
Example of a configuration with 4 peaks of the
PARPM for L = 18. Depending on the position where the tilted tiles reach 
the interface, several distinct processes occur (see the text).}
\label{fig2}
\end{figure}

  A Dyck path is seen as an interface separating a film of 
tilted tiles deposited
on a substrate from a rarefied gas of tiles (see Fig.~\ref{fig2}). The stochastic processes in discrete time  has
two steps:
\vspace{0.5cm}

  {\bf a) Sequential updating}. With a probability $P(i)$ a tile hits the site $i =
1,\ldots, L - 1$
($\sum_iP(i) = 1$). In the RPM, $P(i)$ is chosen uniform: $P(i) = P = 1/(L-1)$.
 In the PARPM, this is not longer the case. For a given configuration $c$ 
(there are $Z(L)$ of them) with $n_c$ peaks. All the peaks are hit with the same probability
$R_p = p/(L-1)$ ($p$ is a non-negative parameter),  all the other $L-1-n_c$ sites are
hit with the same  probability $Q_c = q_c/(L-1)$. Since 
\be \label{e2}
  n_c R_p + (L-1 - n_c)Q_c = 1          
\ee
$q_c$ depends on the 
configuration $c$  and on the parameter $p$, and we have that
\be \label{e1}
  q_c = (L-1-pn_c)/(L-1-n_c), \quad      c = 1, 2,\ldots,Z(L).       
\ee 

\vspace{0.5cm}
  {\bf b) Effects of a hit by a tile}. The consequence of the hit 
on a
configuration is the same as in the RPM at the conformal invariant point.
Depending of the slope $s_i=(h_{i+1}-h_{i-1})/2$
at the site $i$, the following processes can occur:

\noindent 1) $s_i = 0$ and $h_i > h_{i-1}$ (tile $b$ in Fig~\ref{fig2}). The tile hits a peak and is reflected.

\noindent 2) $s_i = 0$ and $h_i < h_{i-1}$ (tile $c$ in Fig.~\ref{fig2}) . The tile hits a local minimum and is  adsorbed ($h_i\rightarrow  h_i + 2$).

\noindent 3) $s_i = 1$ (tile $a$ in fig.~\ref{fig2}). The tile is reflected after triggering the desorption ($h_j \rightarrow h_j-2$) of a layer of $b-1$ tiles from the segment $\{j=i+1,\ldots,i+b-1\}$ where $h_j>h_i=h_{i+b}$. 

\noindent  4) $s_i = -1$ (tile $d$ in Fig.~\ref{fig2}).  The tile is reflected after
triggering the desorption ($h_j \rightarrow h_j-2$) of a layer of
$b-1$ tiles belonging to the segment $\{j=i-b+1,\ldots,i-1\}$
where $h_j>h_i=h_{i-b}$.

 The continuous time evolution of a system composed by the states
$a = 1,2,\ldots,Z(L)$
with probabilities $P_a(t)$ is given by a 
master equation that can be interpreted
as an imaginary time Schr\"odinger equation:
\begin{equation}\label{e3}
\frac{d}{dt} P_a(t) = -\sum_b H_{a,b} P_b(t).
\end{equation}
 The Hamiltonian $H$ is an $Z(L)\times Z(L)$ intensity matrix: $H_{a,b}$ 
($a\neq b$) is non positive  
 and $\sum_a H_{a,b} = 0$. $-H_{a,b}$ ($a\neq b$) is the rate for the transition
$\ket b \rightarrow \ket a$. The ground-state wavefunction of the system $\ket0$, $H \ket0 = 0$, gives
the probabilities in the stationary state:
\begin{equation} \label{e4p}
\ket0 = \sum_a P_a \ket a,\;\;\;\;\;\; P_a = \lim_{t \to \infty}  P_a(t). 
\end{equation}

 In order to go from the discrete time description of the stochastic model
to the continuous time limit, we take $\Delta t = 1/(L-1)$ and
\be \label{e5p}
H_{ac}= - r_{ac} q_c \quad (c\neq a),
\ee
where $r_{ac}$ are the rates of the RPM and $q_c$ is given by Eq.~\rf{e1}. The
probabilities $R_p$ don't enter in \rf{e3}  since in the RPM when a tile hits
a peak, the tile is reflected and the configuration stays unchanged.
Notice that through the $q_c$'s the matrix elements of the Hamiltonian
depend on the size of the system and the numbers of peaks $n_c$ of the
configurations.

As can be seen from (2.3) and (2.7), for $p<1$ the adsorption and desorption 
are faster than at $p=1$ and slower for $p>1$. The  slowing down is 
extreme for the substrate where $n_c=L/2$. In this case for the value 
$p=\pmax=2(L-1)/L$ we have $q_c=0$ and the substrate becomes an absorbing state.

In a previous paper it was shown that the PARPM is conformal invariant in 
the domain $0\leq p<\pmax$. 
 The following exact results, which are independent on $p$, are known for this 
domain. 

 The average height for large values of the size $L$ of the system is equal 
to \cite{JS,AR2} 
\be \label{2.6}
h(L) = \frac{2}{\pi} \int_{\frac{\pi}{L}}^{\frac{\pi}{2}} \frac{\sqrt{3}}{2\pi} 
\ln(\frac{L}{\pi}\sin x) dx +\beta \approx 0.1056 \ln{L} +\beta',
\ee
where $\beta$ and $\beta'$ are non universal numbers. 
 
The density of contact points $g(x,L)$ ($x$ is the distance to the origin), 
 in the finite-size scaling limit ($x >> 1$, $L >> 1$, but $x/L$ fixed) is 
given by \cite{APR}:
\be \label{2.7}
 g(x,L) =  C\left( \frac{L}{\pi}\sin(\pi x/L)\right)^{-1/3},
\ee
where 
\be \label{2.8}
 C = -\frac{\sqrt{3}}{6\pi^{5/6}} \Gamma(-1/6) = 0.753149... \quad .
\ee
 The average density of minima and maxima (sites where adsorption doesn't take 
place) $\tau$(L) has the asymptotic 
value:
\be \label{2.9}
 \lim_{L \to \infty} \tau(L) = 3/4,
\ee
with non-universal corrections (depending on the value of $p$) of order $1/L$.
 We will use these results in Sections 3 and 6. 
 
\section{  Quasistationary states at $\pmax$}

 When we understood that at $\pmax$ one has an absorbing state, and
therefore a phase transition, we got interested to see how conformal
invariance is broken. We expected the system to get massive as is the
usual case when conformal invariance is broken. It turns out that the new
phase is a fascinating object.

\begin{figure}
\centering
\includegraphics[angle=0,width=0.5\textwidth] {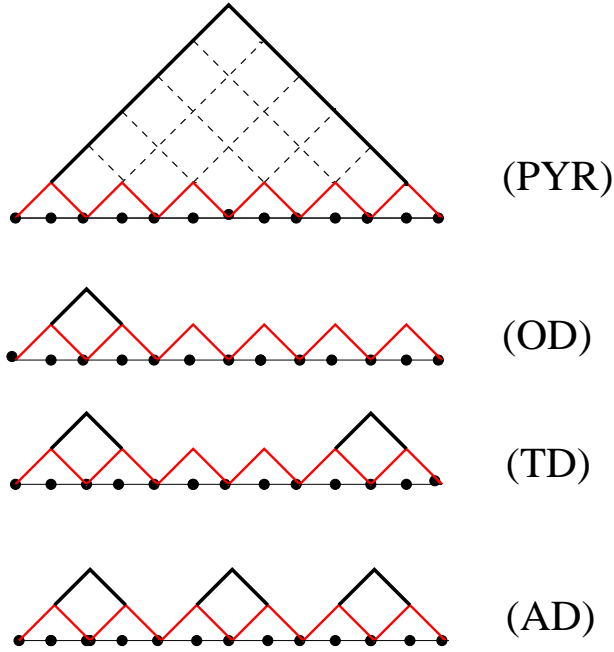}
\caption{
Special initial conditions for $L= 12$ (see text).}
\label{fig3}
\end{figure}

 We have studied several initial conditions specified by the
local heights $h_i$ (see Fig.~\ref{fig3}). One is the "{\it pyramid}" (PYR): the heights are: $0,1,2,
3...,L/2, L/2 -1,...,1,0$. Another is the "{\it one dent}" (OD): the heights
being $0,1,2,1,0,1,0,1,0,...,1,0$. The "{\it two dent}" 
(TD) one: the heights are:
$0,1,2,1,0,1,0,...,0,1,2,1,0$. The "all dents" (AD) is defined by the local
heights: $0,1,2,1,0,1,2,1,0,...,0,1,2,1,0$.
We have chosen these four
configurations because they are extreme cases. The PYR configuration has
only one peak. The OD asymmetric configuration has only one peak less than
the substrate which has the maximum number of peaks. The symmetric TD
configuration has two peaks less than substrate, the AD configuration is
an intermediate one.

\begin{figure}
\centering
\includegraphics[angle=0,width=0.5\textwidth] {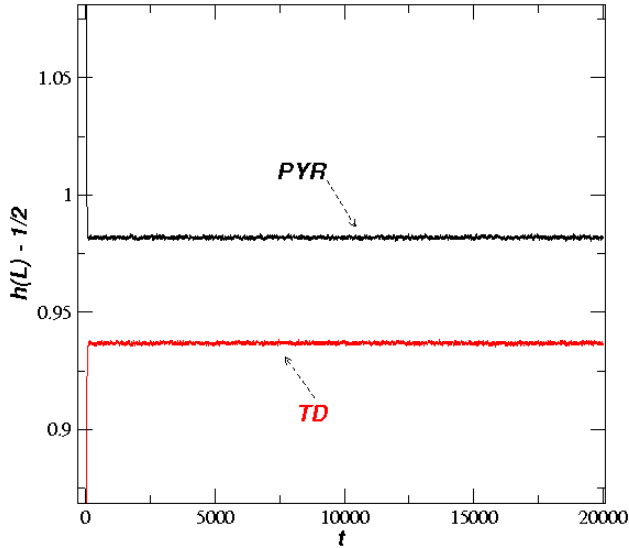}
\caption{
 $p=\pmax$. The average height from which one has subtracted the average
height of the substrate ($\frac{1}{2}$), as a function of time for two initial conditions
PYR and TD. The lattice size $L = 96$. The averages are obtained taking
$6\times10^5$ samples.}
\label{fig4}
\end{figure}

 We have looked, in Monte Carlo simulations, at the average height from
which we have subtracted the average height of the substrate (equal to
1/2), as a function of time taking $L = 96$, and starting with the PYR and TD
initial conditions. The results of our simulations are shown in Fig.~\ref{fig4}. 
 One
sees that for each of the two initial conditions one obtains time-independent
results which are not zero, as one could expect to find in the absorbing
state. This observation suggests the existence of {\it quasistationary states}
\cite{CDR,DMU}.

\begin{figure}
\centering
\includegraphics[angle=0,width=0.5\textwidth] {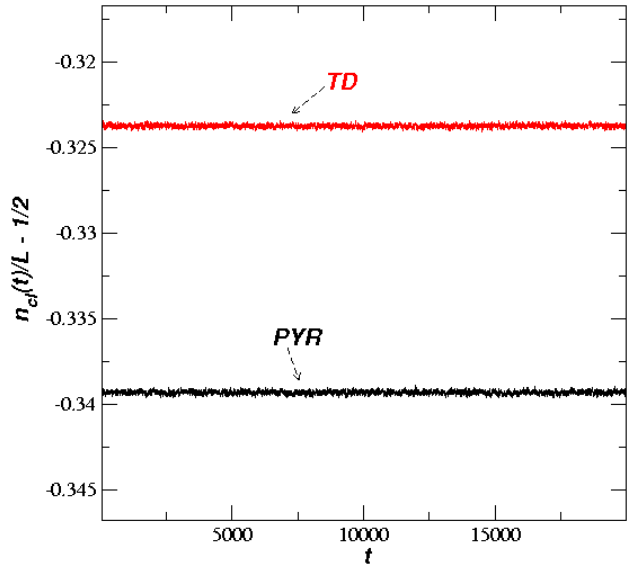}
\caption{
 The density of clusters - 1/2 (the density of clusters in the
substrate) as a function of time. Same conditions as in Figure 3.
}
\label{fig5}
\end{figure}

 We have also looked at the average density of clusters $n_{cl}(t)/L$ from
which we have subtracted $\frac{1}{2}$ which is the corresponding quantity for the substrate.
 The results for the two initial conditions PYR and TD are
shown in Fig.~\ref{fig5}. One obtains similar results as  those shown in Figure  \ref{fig4}.

 Comparing the average heights and the density of clusters for the two
initial conditions, suggest that one has more clusters (therefore lower
values of the average height) for the TD initial condition compared to the
PYR initial condition. Therefore for $L = 96$ the two QSS are different.

 Let us make an observation.  If we assume that in the QSS one has the same
density of clusters as in the stationary distribution observed in the
$0\leq p < \pmax$ case \cite{GNP}, for which one has an analytical expression ($n_{cl}/L
= \Gamma(1/3)\sqrt{3}/2\pi L^{1/3}$), one obtains the value $-0.339$ for the
quantity shown in Fig.~\ref{fig5}. This value is closed to   the value seen for the PYR initial
condition. This observation will play in important role in understanding
the QSS.

 In order to see how the QSS appeared, we looked for another
quantity which was extensively studied in the PARPM \cite{AP}. This is the
average density of sites where one has a maximum or a minimum in a given
configuration
  $\taumax$ for $p=\pmax$. In the PARPM and $0 \leq p < \pmax$, in the large $L$
limit one has $\tau(L,\infty)=\tau (L) = 0.75$ (for the substrate $\tau(L) = 1$).

\begin{figure}
\centering
\includegraphics[angle=0,width=0.5\textwidth] {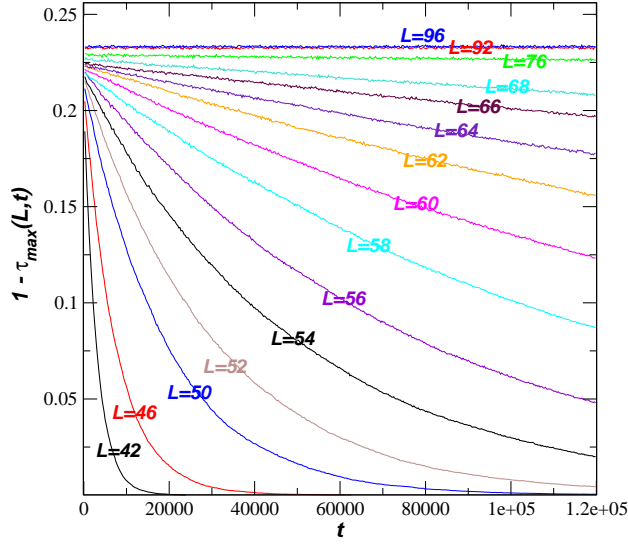}
\caption{
 $p=\pmax$. Average density of minima and 
maxima $\taumax$ subtracted from 1, as a function of time for different lattices $L$. The initial
condition OD was chosen. The averages are obtained by taking $6\times10^5$ samples.
}
\label{fig6}
\end{figure}

\begin{figure}
\centering
\includegraphics[angle=0,width=0.5\textwidth] {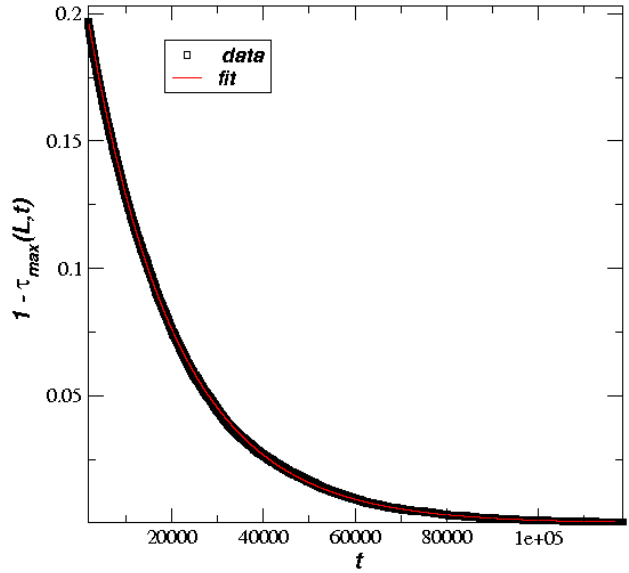}
\caption{
 The average density of minima and maxima $\tau (L)$ as a
function of time for $L = 50$ together with the fit (3.2). 
The initial configuration is OD and the averages are obtained by 
taking $6\times10^5$ samples.
}
\label{fig7}
\end{figure}

 If one starts with the TD configuration and looks at time variation of
the quantity $1- \taumax $ for different system sizes one obtains the
results shown in Fig.~\ref{fig6}.
 We can see that for small values of $L=46 - 60$ one has an
exponential fall-off with time. Then an almost linear decrease in time
 for the values $L = 62-76$. For $L = 92$ and $96$ one sees
practically no time variation and a value $\taumax \sim 0.77$ very close to
the value $0.75$ seen in the stationary state in the conformal invariant
domain of the model.
 This makes us suspect the following behavior of $\taumax$ in the QSS:
\be \label{3.1}
  1- \taumax = A(L) \exp(-E(L)t),
\ee
 where $E(L)$ decreases exponentially with $L$ and $A(L)$ increases smoothly
with $L$ to the value $0.25$ observed for the stationary states in the 
$0\leq p <\pmax$ domain. In Fig.~\ref{fig7}, we show for $L = 50$ the data 
undistinguished from a fit:
\be \label{3.2}
1-\tau_{\mbox{\scriptsize{max}}}(50,t) = 0.217 \exp (-0.0000522t).
\ee
 Notice the very small value of $E$ and the fact that $A(50)$ is not far away
from the value $0.25$.

 We sum up our observations; although for $p = \pmax$ we expected the system
to relax in the absorbing state, we observed the existence of states with
very long relaxation times. For the lattice sizes presented above, the
QSS depend slightly on the initial conditions. Surprisingly, the
density of clusters and of maxima and minima in the QSS have values closed
to those observed in the stationary states for $0 \leq p < \pmax$. Actually for
the PYR initial condition the results coincide.
 In the next three Sections we will explain these observations.

\section{ The spectrum and wavefunctions of the Hamiltonian. How QSS occur at 
$\pmax$.}

 In order to understand the origin of the QSS we have studied the spectrum 
of the Hamiltonian which gives the time evolution of the system (see Eq. 
(2.5)). For $0\leq p < \pmax$, where we have conformal invariance, we have 
checked \cite{AP} that in the finite-size scaling limit:
\be \label{4.1}
\lim_{L\to \infty}E_i(L) = \pi v_s\Delta_i/L, \quad i=0, 1, 2, \ldots,
\ee
where $E_0 = 0$, $\Delta_i$ are the scaling dimensions, and the sound 
velocity has the expression:
\be \label{4.2}
v_s(p) = (1 - 3(p - 1)/5) 3\sqrt{3}/2.
\ee
Notice that the velocity decreases when $p > 1$ since, as described in 
Section 2, the transition rates are smaller. The scaling dimensions are 
given by the partition function \cite{SSS}:
\be \label{4.3}
Z(q) = \sum_{i=0}^{\infty} q^{\Delta_i} = (1 - q) \prod_{n=1}^{\infty}(1 - q^n)^{-1}.
\ee
We give the first values of $\Delta_i$'s together with the corresponding 
degeneracies ($d_i$'s):
\be \label{4.4}
 \Delta = 0 (1), 2 (1), 3 (1), 4 (2), 5 (2), 6 (4), 7 (4),...     .
\ee
 We will check if these values will be seen also for $p = \pmax$. 

\begin{figure}
\centering
\includegraphics[angle=0,width=0.5\textwidth] {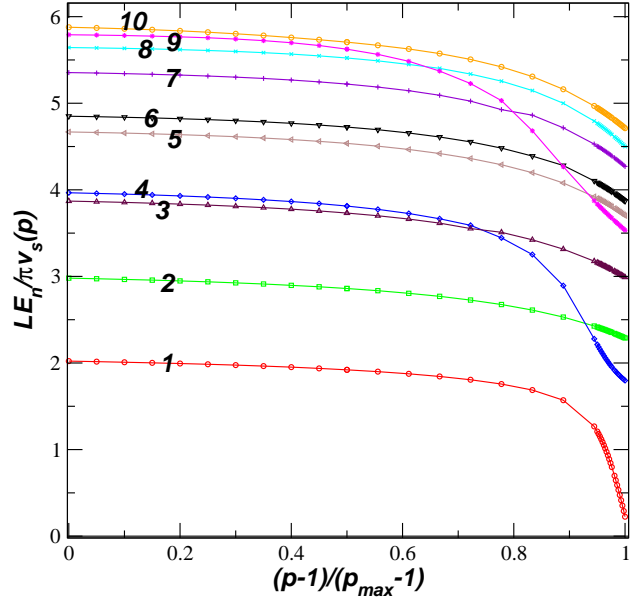}
\caption{
 Estimates $\Delta_n^{(n)} = LE_n/\pi v_s(p)$ of the scaling dimensions 
$\Delta_n$ for different 
values of $p$, and for the lattice size $L=18$.
 The estimates corresponding to the first 11 energy levels
 are shown. The values of $v_s(p)$ were obtained by (4.2).}
\label{fig8}
\end{figure}

 In order to estimate the values of the $\Delta_i$'s, we have taken $L = 18$ 
(this is not a small lattice!) and diagonalized numerically the Hamiltonian for 
various values of $p$. The results are shown in Fig.~\ref{fig8} where the first 11 
levels are seen (the ground-state energy $E_0$ is equal to zero). The 
remaining 10 levels should correspond roughly (see (4.4)) to $\Delta = 
2,3,4,5$ and $6$. We can see that for $p = 1$ where the model is integrable, 
this is indeed the case. The levels cluster in the right places. When $p$ 
increases, one notices that the properly scaled $E_1$ after a smooth 
behavior up
to $p \approx 0.9$, decreases rapidly for $p = \pmax$ (we have  used for
$v_s(\pmax)$ the value given by (4.2) for $p = 2$). 
Using Monte Carlo simulations,
we have checked \cite{AP} that for $p < \pmax$ the small decrease with $p$ of $E_1$
 is 
a finite-size effect and therefore what one sees in the figure is a 
crossover effect. One can also see that for increased values of $p$, $E_4$ 
crosses $E_3$ and that $E_9$ crosses all the levels $E_8,\ldots,E_5$. 
Except for 
the three levels $E_1, E_4$ and $E_9$ which decrease dramatically for 
$p = \pmax$, the other levels have the same finite-size behavior as 
those in the
conformal invariant domain ($p < \pmax$). This suggest that the three levels
mentioned above might be related to QSS. We now proceed to a detailed analysis
of these observations.  
 
\begin{figure}
\centering
\includegraphics[angle=0,width=0.5\textwidth] {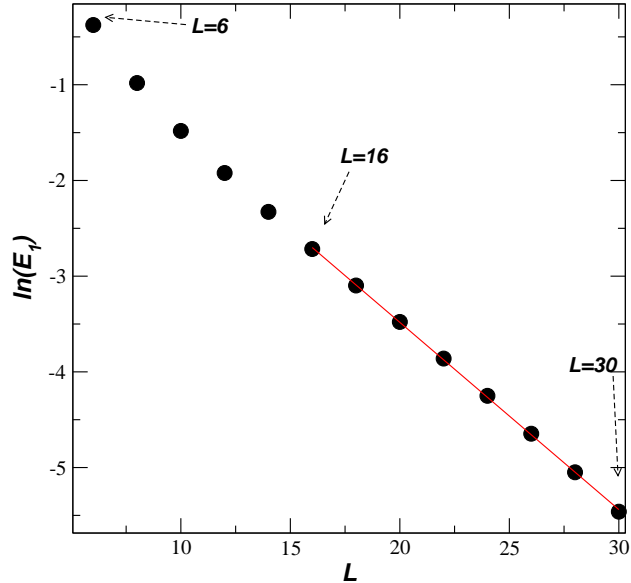}
\caption{
 $\ln(E_1)$ as a function of the lattice size $L$ for $p = \pmax$.
The red line is a guide for the eyes, and is obtained from a fit
 where the lattice sizes 
$L=16-30$ were used.}
\label{fig9}
\end{figure}

 Using different lattice sizes we have computed $E_1$ as a function of $L$ up 
to $L = 30$. For this calculation we could study larger lattices  
due to a special property of the Hamiltonian at $p=\pmax$. As we are going to show in section 5, the eigenlevel corresponding 
to $E_1$ is the ground state energy of a reduced matrix defined in a basis 
where the absorbing state is absent. In this case, by using the power 
method we were able to calculate $E_1$ up to $L=30$.  The
results can be seen in Fig.~\ref{fig9}. Using the two points 
corresponding to the largest lattice sizes, one obtains:
\be \label{4.5}
 E_1 (L) = 0.912 \exp(-0.206L).
\ee

\begin{figure}
\centering
\includegraphics[angle=0,width=0.5\textwidth] {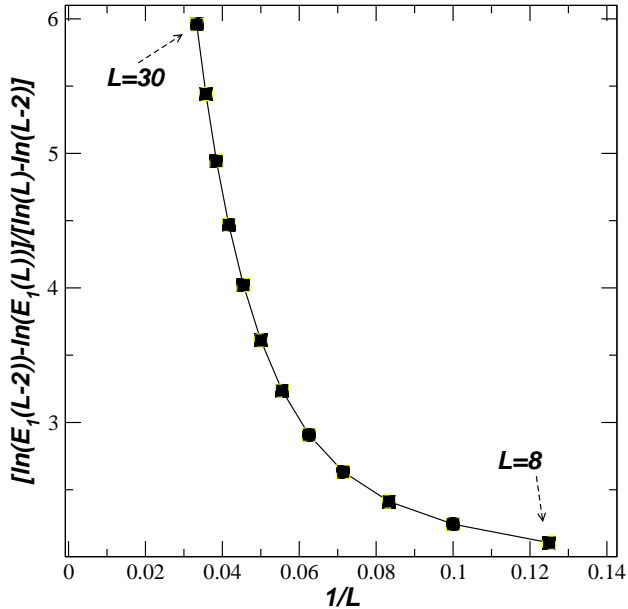}
\caption{
 Estimates of $-d/\ln(L)\{\ln(E_1(L))\}$ as a function of $1/L$ for 
different lattice sizes ($p = \pmax$).}
\label{fig10}
\end{figure}

 To be sure that a power law behavior ($E\approx L^{-m}$) is excluded, 
we have 
estimated the derivative $-d/d\ln(L)\{\ln(E_1(L))\}$. 
This quantity should reach a 
constant for large values of $L$ if $E_1(L)$ is power behaved but 
should diverge linearly in the case of an exponential behavior like (4.5).
As seen from Fig.~\ref{fig10}, the exponential behavior (4.5) is correct.  

 A similar analysis of $E_4(L)$ but up to $L = 18$ only, gives a similar 
result:
\be \label{4.6}
  E_4 (L) = 2.41 \exp(-0.10L). 
\ee
 We have not looked at $E_9(L)$ but we expect again 
an exponential fall-off. 
We conclude that three energy levels have an exponential fall-off and that 
they can be related to QSS. This is going to be shown to be the case in 
Section 5. We proceed by looking at the remaining levels.

 The data suggest that at $p = \pmax$, the energy levels 
 that do not go exponentially to zero have the same 
finite-size scaling behavior as those in the conformal invariant domain. 
This would imply that instead of (4.4) we would have
\be \label{4.7}
 \Delta = 0 (1), 3 (1), 4(1), 5 (2), 6 (3), \ldots \quad .
\ee
 If confirmed, this would lead us to a strange picture since the scaling 
dimension $\Delta = 2$ does not appear. This dimension corresponds 
to the energy-momentum, 
 and therefore conformal invariance couldn't apply and  we 
could not explain the finite-size behavior of the remaining levels. What 
can go wrong in our picture? One possibility is that the finite-size 
scaling of the levels doesn't satisfy (4.1).  

\begin{figure}
\centering
\includegraphics[angle=0,width=0.5\textwidth] {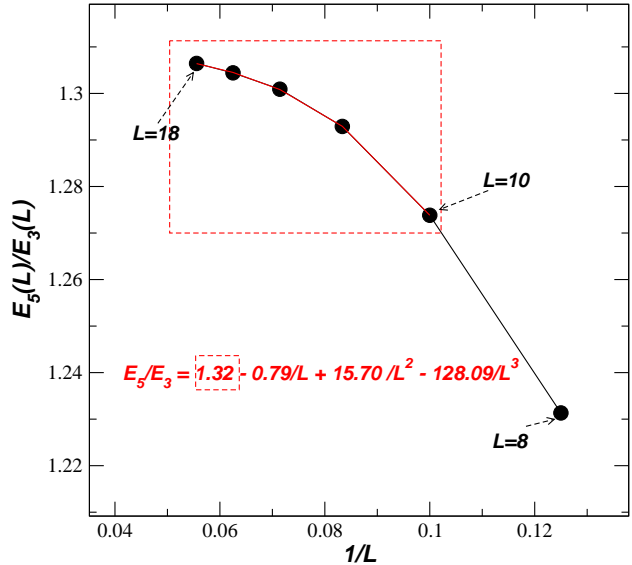}
\caption{
 Ratio among the two lowest eigenenergies, at $p=\pmax$,that does not vanish exponentially. The data are plotted as a 
 function of $1/L$, for lattice sizes $L=8-18$. A cubic fit was done by using the five largest lattice sizes.}
\label{fig11}
\end{figure}

\begin{figure}
\centering
\includegraphics[angle=0,width=0.5\textwidth] {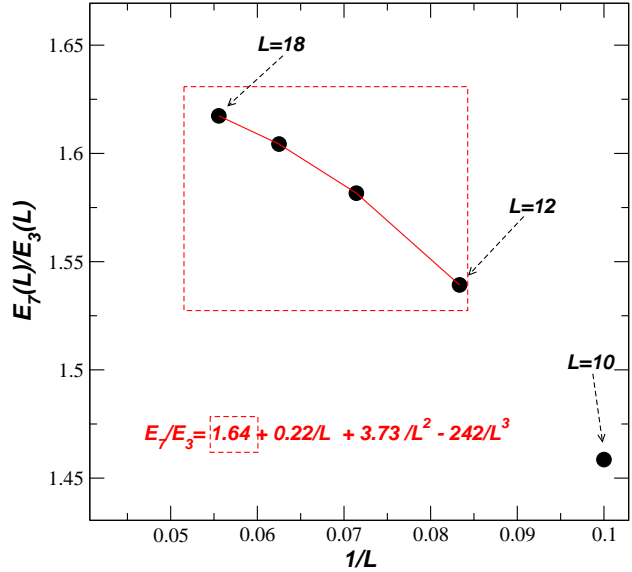}
\caption{
	Ratio among the third and second lowest eigenenergies, at $p=\pmax$,that does not vanish exponentially. The data are plotted as a 
 function of $1/L$, for lattice sizes $L=10-18$. A cubic fit was done by using the four largest lattice sizes.}
\label{fig12}
\end{figure}

 We have computed $E_2 (L)$ up to $L = 18$. A fit to the data gives $\Delta_3 
= 3.05$ in agreement with what should be expected. 
Similarly examining $E_4 (L)$ 
 one finds $\Delta_4 = 3.95$ also as expected. These estimates were found 
assuming that $v_s(\pmax)$ is given by Eq. (4.2). Can we get $\Delta = 2$  
changing the sound velocity such that $E_3(L)$ gives $\Delta = 2$, $E_5(L)$ 
gives $\Delta = 3$, $E_6 (L)$ gives $\Delta = 4, \ldots$? We have computed the 
ratios $E_5(L)/E_3(L)$ and $E_6(L)/E_3(L)$ as a function of $L$. 
One should 
obtain 4/3 respectively 5/3 if one had (4.3) and 3/2 respectively 2 if the 
energy momentum tensor would be present. In Figs.~\ref{fig11} and \ref{fig12} 
we show these ratios
as functions of $1/L$. Cubic fits give the values 1.32 respectively 
1.64. We conclude that (4.7) is most probably correct. We have also 
checked that there are no energy crossings for the levels which cluster around
$\Delta = 7$.
 
 The analysis of the energy levels (compare (4.4) with (4.7)) suggests 
that at $p = \pmax$ the partition function (4.3) changes in the following 
way: the degeneracy at each even value of $\Delta$ decreases with one unit. 
Each energy level which left the Virasoro representation at non-zero even 
values of $\Delta$ moves to $\Delta$ = 0 which becomes infinitely 
degenerate. This opens a problem in the representation theory of the 
algebra which might be solvable since the central charge is c = 0. In 
Section 6 we are going to learn more about this puzzle.   

\section{
 From eigenfunctions of the Hamiltonian to QSS}

 We have seen in the last section that some eigenvalues of the Hamiltonian 
vanish exponentially. Here we will show what is their connection to QSS. 
In order to do so, we first prove a special property of Hamiltonians  in the
presence of an absorbing state.

 We denote the vector space corresponding to $n + 1$ configurations by 
$\ket{0}$, 
$\ket{i}$ ($i = 1,2,...,n$), in which we have chosen $\ket{0}$ to be the absorbing state.
The Hamiltonian has the following properties:
\be \label{5.1}
 H_{i,0} = 0,\quad  H_{0,0} = 0, \quad H_{i,j} \leq 0,
\ee    
\be \label{5.2}
  H_{0,j} + \sum_i  H_{i,j} = 0 \quad (j = 1,2,...n).
\ee
 Assume that $E_k$ ($k = 1,..., n$) is a non-vanishing eigenvalue of $H$ 
with an eigenvector ($y_0^{(k)}, y_1^{(k)},...,y_n^{(k)}$). 
We can show that the 
sum of the components of any eigenvector is equal to zero: 
\be \label{5.3}
y_0^{(k)} + \sum_i y_i^{(k)} = 0 \quad (k = 1,2,..,n).
\ee
 Using \rf{5.1} and \rf{5.2} we have:
\be \label{5.4}
E_k y_0^{(k)} = \sum_iH_{0,i}y_i^{(k)} = - \sum_{j,i} H_{j,i} y_i^{(k)} = - 
E_k \sum_j y_j^{(k)},       
\ee
from which the identity (5.3) follows.

 We consider now the reduced matrix $H'_{i,j}=H_{i,j}$ ($i,j = 1,2,...,n$) 
(the configuration $\ket{0}$  
is taken out). Let $E_1$ be the lowest non-vanishing eigenvalue, from \rf{5.1} 
and using the Perron Frobenius theorem, we get:
\be \label{5.5}
y_i^{(1)} \geq  0,
\ee          
and using (5.3), $y_0^{(1)}<0$. 
From the same theorem we also learn that $E_1$ is the unique eigenvalue 
for which \rf{5.5} occurs. For the other eigenvalues, at least one component 
of the wavefunction is negative, i. e. Eq.~(5.5) is not valid for 
$y_i^{(k)}$ ($k>1$).

 The solutions of the differential equations \rf{e3} are
\ba \label{5.6}
&&P_0(t) = 1 + \sum_k A_k y_0^{(k)} \exp(-E_k t), \nonumber \\
&&P_i(t) = \sum_k A_k y_i^{(k)} \exp(-E_k t).      
\ea
The $n$ constants $A_k$ are determined from the initial conditions.

At large values of $L$ and $t$, the exponentially falling energies
give the major contributions to $P_0(t)$ and $P_i(t)$. Among them, $E_1$
 plays a
special role. Note only is $E_1$ the smallest energy, but the components of
its eigenfunction are positive (5.5). This implies that for a large range
of $L$ and $t$ (both of them large), one can keep only the  term with 
$k = 1$ in the
sums appearing in (5.6). The situation is different at very large values
of $L$ when all  the exponentially falling energies are practically equal
to zero and more terms can appear in (5.6). Independent of the initial
conditions, the term with $k = 1$ has to be present in the sums (5.6) in 
order to
 assure the positivity of the probabilities $P_0(t)$ and $P_i(t)$
 since for the
other eigenfunctions (5.5) is not valid. For example, the eigenfunctions
corresponding to the levels $E_4 (L)$ and $E_9 (L)$ of Fig.~\ref{fig8} have components
with both signs in the sum giving $P_i(t)$.

 We are now in the position to show how exponentially falling energies 
like $E_1(L)$ (see \rf{4.5}) can be the origin of QSS. 
For our discussion we
will assume that the term with $k = 1$ alone is present in (5.6). It turns
out that this assumption will help to explain all the results obtained for
the Monte Carlo simulations. The probability vector $\ket{P(t,L)}$ becomes:
\be \label{5.7}
 \ket{P(t,L)} = [1 - a(L)\exp(-E_1(L)t)]\ket{0} + a(L) \sum_i p_i(L) 
\ket{i}\exp(-E_1(L)t),   
\ee
where we have used \rf{5.3} and \rf{5.5} 
\be \label{5.8}
a(L) = A_1\sum_i y_i^{(1)}(L); \quad  p_i(L) = y_i^{(1)}(L)/\sum_i 
y_i^{(1)}(L).
\ee
 Note that $p_i(L)$ gives the probability to find the system in 
configuration $\ket{i}$ if the system is in the stationary state of the 
reduced Hamiltonian $H'$, that acts in the vector space where the substrate 
is absent. 
 Thus, Equation \rf{5.7}  explains the occurrence of QSS. If $L$ is large enough, 
$E_1(L)$ is negligible, one finds:
\be \label{5.9}
\ket{P(L)} = (1-a(L))\ket{0} + a(L) \ket{P_{\mbox{\scriptsize ns}}(L)}.         
\ee
 Here $a(L)$ depends on the initial conditions and it is not equal to zero.
 $\ket{P_{\mbox{\scriptsize ns}}(L)}$ is a probability distribution function 
of configurations in which the substrate is not present.  
Equation \rf{5.9} describes therefore the QSS. 
 Using Monte Carlo simulations we have measured the probability to find 
the system in the substrate. For the OD initial condition (see Section 
3 for the definition), one finds:
\be \label{5/10}
1-a(L) \sim  4.5/L.
\ee 
For the PYR initial condition one finds $1 - a(L) = 10^{-5}$ for $L = 96$ 
(we didn't compute $a(L)$ for the TD and AD initial conditions). 
This implies that for large values of $L$ one has 
\be \label{5.11}
\ket{P(L)} = \ket{P_{\mbox{\scriptsize ns}}(L)}, 
\ee
and the substrate does not occur in the QSS. We postpone the discussion of the
results shown in Fig.~\ref{fig6} up to the next Section.

\section{ The quasistationary states}

 In this section we are going to identify the quasistationary states 
observed in the Monte Carlo simulations (see Figs.~\ref{fig4}-\ref{fig7}) and 
 find a puzzle. The basis of our identification are Eqs.~(5.7), (5.9) and
(5.11).

\begin{figure}
\centering
\includegraphics[angle=0,width=0.5\textwidth] {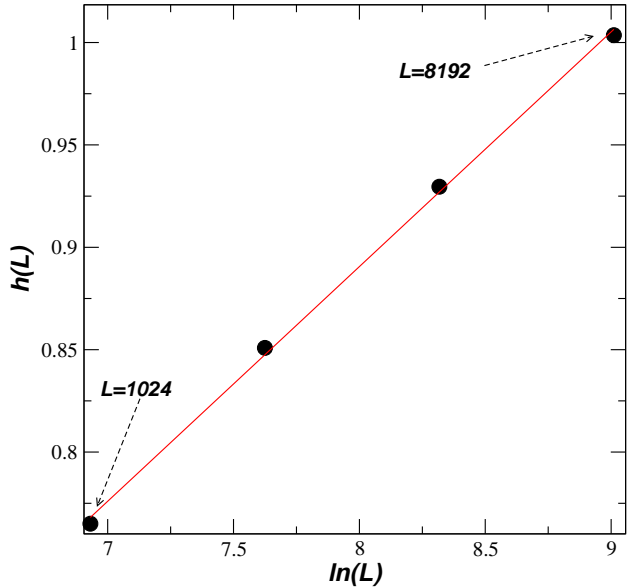}
\caption{
 The average height $h(L)$ as a function of $\ln(L)$ in the QSS. $L = 
1024, 2048, 4096$ and $8192$. We have used PYR as an initial distribution.}
\label{fig13}
\end{figure}

 We have taken large lattices and started our simulations with the PYR 
initial condition. We first looked at the average height $h(L)$. The results 
are shown in Fig.~\ref{fig13}. The data can be fitted by a straight line. Taking only the two largest lattice sizes we obtain:
\be \label{6.1}
h(L) = 0.1068 \ln(L) + 0.042.
\ee    
In astonishing agreement with the expression (2.8) derived \cite{JS,AR2} assuming 
conformal invariance and used to describe the data for $0\leq p < \pmax$. Since 
for the large lattices we have considered, we can use (5.11) and this 
would imply that $\ket{P_{\mbox{\scriptsize ns}}(L)}$ is given by the same function as 
the one seen away
from $\pmax$. 

 If confirmed, this result would be surprising because as we have 
discussed in Section 4, as compared to the conformal invariant domain, at 
$p = \pmax$ the finite-size spectrum of the Hamiltonian is a "mutilated" 
one. It lacks not only the energy momentum tensor but other levels too. On 
the other hand that space-like correlation functions look to be 
unaffected.

\begin{figure}
\centering
\includegraphics[angle=0,width=0.5\textwidth] {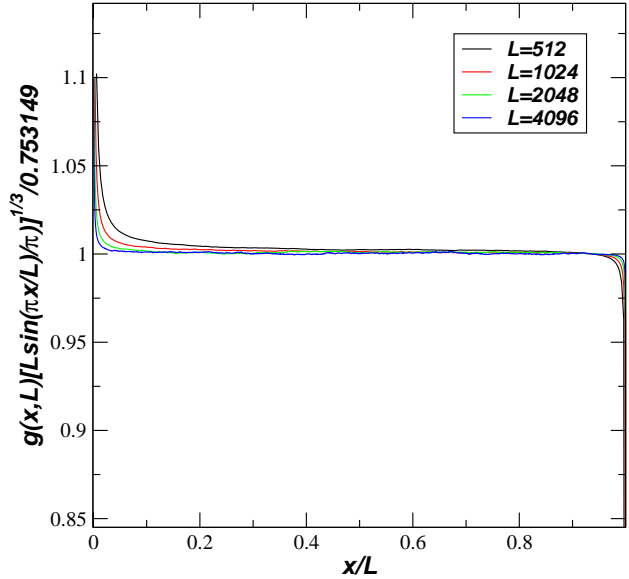}
\caption{
 The density of contact points $g(x,L)$ as a function of $x/L$ at 
$\pmax$, divided by (2.9). $L = 1024, 2048, 4096$ and $8192$. Initial condition 
AD.
The average are obtained from 3500 samples running $10^6$ mcsteps. }
\label{fig14}
\end{figure}

 In Fig.~\ref{fig14} we show  for the AD initial condition, the density of contact points $g(x,L)$ for various 
lattice sizes divided by the finite-size scaling distribution (2.9) in the 
QSS. The 
coincidence of the $\pmax$ data and the expectation coming from conformal 
invariance in the QSS  is astonishing
(for the large lattices considered here $a(L)$ is
practically equal to 1).

 From now on, we will assume that for large lattices  the QSS the correlators
are those of the conformal invariant model (RPM) and will try to explain 
the results described in Section 3.

 We use Eq.~(5.9) and the observation that for the initial condition PYR 
$a(L) = 1$,  taking $a(96) = 0.975$ for the TD initial condition we obtain 
the results for TD presented in both Figs.~\ref{fig4} and \ref{fig5}
(we don't have at
hand an independent measurement of $a(L)$ for the TD initial condition).  

\begin{figure}
\centering
\includegraphics[angle=0,width=0.5\textwidth] {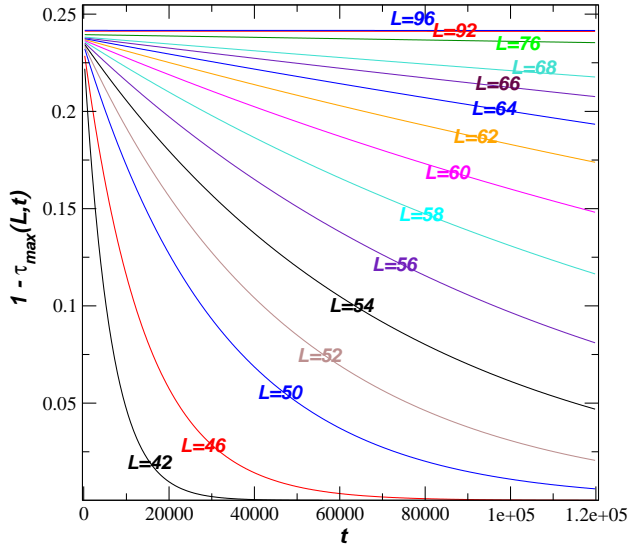}
\caption{
The average density of minima and maxima $\taumax(L,t)$ subtracted from 1, as a function of time for the same lattice sizes $L$, as in Fig~\ref{fig6}. The initial 
condition is OD. The data were generated from the prediction (6.3), using 
$E_1$ given by (4.5)}
\label{fig15}
\end{figure}

We proceed by looking for a 
  description of the data presented in  Fig.~\ref{fig6}, where the time 
dependence of the quantity $1 - \taumax $, taking OD initial condition, is shown.  $\taumax$ is the 
average density of minima and maxima which is equal to 1 for the 
substrate. We use Eqs.~(5.7) and (5.10)  to get:
\be \label{6.2}
1 - \taumax = [-4.5/L + 1 - (1 - 
4.5/L)\tau(L)]\exp(-E_1(L)t).                  
\ee
The function $\tau(L)$ is known analytically only for $p = 1$ \cite{GNP}. It is
equal to $3/4$ in the large $L$ limit and has non-universal corrections of
order $1/L$ which are unknown for $p = \pmax$. In principle these corrections
can be determined by looking at the QSS obtained using PYR initial
conditions.
 We can however use the result obtained in \cite{AP} for $p = 1.99$
where we found $\tau(L) \approx 0.75 - 0.3/L$ and use it in (6.2) to get a fair
description of the data. We obtain

\be \label{6/3}
 1 -\taumax  \approx  (0.25 - 0.8/L) \exp(-E_1(L)t).
\ee                
The value of $E_1(L)$ can be estimated from (4.5).
 In Fig.~\ref{fig15} we use this function, together with the predicted values of $E_1(L)$ given in (4.5), to compute the time dependence of $1 - \taumax$ for the same lattice sizes used in Fig.~\ref{fig6}. We can see that the overall behavior of Fig.~\ref{fig15} and Fig.~\ref{fig6}, generated by the Monte carlo simulations,  are the same.

\section{Conclusions}

 The parameter $p$ which enters in the definition of peak adjusted raise and 
peel stochastic model determines a domain $0\leq p < \pmax$ in which one has
conformal invariance. In the finite-size scaling regime all the properties are 
independent on $p$ which appears only in the sound velocity $v_s(p)$ which 
fixes the time scale. If $p > 1$, global properties of the configurations 
(the number of peaks) and the size of the system $L$, enter in the rates. 
The larger is the number of peaks of a configuration, the smaller are the
rates to escape the configuration. As a result, at $p = \pmax = 2(L-1)/L$ 
the configuration with the largest number of peaks which is the substrate, 
becomes an absorbtive state for any size of the system. Since there are 
no fluctuations in the ground-state of the system, we expect conformal 
invariance to be lost and get into a massive phase. This is not the case 
and a fascinating phenomenon takes place.

 The new features of the model at $p = \pmax$ are encoded in Figures \ref{fig4} and \ref{fig6}. 
If the evolution of the system starts with a given initial configuration,
instead of moving fast to the absorbing state, the system gets stuck in 
another configuration and the relaxation time grows exponentially with the 
size of the system. This is a quasistationary state. In Figure \ref{fig8} we look 
at the spectrum of the Hamiltonian and follow the change of the scaled
energies with increasing values of the parameter $p$. From the eleven energy
levels shown in the figure, eight of them are those of the conformal invariant
region ($0\leq p < \pmax$). Three of them go exponentially to zero for large
values of $L$. 

 In this paper we have tried to clarify this phenomenon. Based, 
unfortunately, on numerics, the following picture emerges. The 
wavefunction corresponding to the first excited state whose energy 
vanishes exponentially,  gives the probability distribution function of the
QSS. This is due to a peculiar property of the first exited state of any 
stochastic Hamiltonian in the presence of an absorbing state. For finite 
values of $L$, the QSS depends on the initial conditions but in the 
thermodynamical limit, it becomes independent of them. Unexpectedly, the 
finite-size scaling properties of the QSS are identical to the one 
observed for the ground-state in the conformal invariant domain.

 This picture describes a strange way to break conformal invariance. Part
of the spectrum stays unchanged (not the scaling dimension of the
energy-momentum tensor!) and another part sinks in the vacuum which
becomes probably infinite degenerate. We have not a clue how to explain
this observation. At the same time, the space-like correlations seen in
the QSS are those of the conformal domain. This is obviously an
unexplained paradox. The study of space-time correlation functions
should shade light in this problem. We plan to do it in the future.

 We have studied only four initial conditions and found essentially only
one QSS. This QSS could be related to the eigenfunction corresponding to
the
first exited state. It is most probable that we have missed other QSS
which should be related to some linear combination of eigenfunctions
corresponding to the remaining exponentially decaying energies.

 The search for QSS should continue in some extensions of the PARPM. One
can look at the fate of defects like those studied in \cite{AR} when the rates
are adjusted to the number of peaks. Another interesting avenue is to
study the effect of changing  the rates depending on the number of peaks
in the extension \cite{GNP,AR3} of the raise and peel model in which the rates of
adsorption and desorption are not equal.

\section{Acknowledgments}

We would like to thank S. Ruffo, H. Hinrichsen, A. Bovier and J. A. Hoyos 
for reading the manuscript 
and discussions.
   This work was supported in part by FAPESP and CNPq 
(Brazilian Agencies).
Part of this work was done while V.R. was visiting the Weizmann 
Institute. V. R. would like to 
thank D. Mukamel for his hospitality and  the support of the Israel 
Science Foundation
(ISF) and the Minerva Foundation with funding from the Federal German 
Ministry for
Education and Research.

\end{document}